\documentstyle[preprint,prl,aps]{revtex}
\narrowtext
\draft

\begin{document}
\title{$s$-wave superconductivity from antiferromagnetic spin-fluctuation model
for
bilayer materials}
\author{A.I.Liechtenstein$^a$, I.I. Mazin$^{a,b}$, and O.K. Andersen$^a$}
\address{$^a$Max-Planck-Institut f\"ur
Festk\"orperforschung,\ Heisenbergstr.1,\ D-70569 Stuttgart, FRG.}
\address{$^b$Geophysical Laboratory, Carnegie Institution of Washington,
5251\ Broad\ Branch Rd., NW, Washington, DC 20015. }
\twocolumn[
\maketitle

\begin{abstract}
It is usually believed that the spin-fluctuation mechanism for
high-temperature superconductivity results in $d$-wave pairing, and that it
is destructive for the conventional phonon-mediated pairing. We show that in
bilayer materials, due to nearly perfect antiferromagnetic spin correlations
between the planes, the stronger instability is with respect to a
superconducting state whose order parameters in the even and odd plane-bands
have opposite signs, while having both two-dimensional $s$-symmetry. The
interaction of electrons with Raman- (infrared-) active phonons enhances
(suppresses) the instability.
\end{abstract}

]
The currently most exciting discussion about high-$T_c$ superconductivity
deals with the symmetry of the pairing state\cite{Nature}. Intimately
related to this, is the question of whether the superconductivity is due to
antiferromagnetic spin fluctuations (see e.g. Monthoux and Pines (MP) Ref.
\cite{MP}, and also Refs. \cite{scal,bickers}), to electron-phonon (EP)
interaction enhanced by inter-layer pair tunnelling\cite{Science}, or to
neither of two. In this discussion, it is indirectly assumed that the
antiferromagnetic spin-fluctuation (AFSF) mechanism necessarily leads to
d-wave pairing, and that the AFSF and EP mechanisms cannot coexist.

In this Letter we point out that, whereas the AFSF mechanism leads to $d$%
-pairing for one layer, it may lead to (two-dimensional) $s$-symmetry for a
bilayer. The condition for that is existence of strong antiferromagnetic
correlations between the two layers in a bilayer, as found experimentally in
YBa$_2$Cu$_3$O$_7$ \cite{tranq,mook,rossat}. We find that, for a given
coupling strength, $T_c(s)$ is about twice as high as $T_c(d)$ thus making
it easier to achieve the observed values of $T_c\sim 100$K. Essential for
the positive influence of layer-doubling is the {\em single}-particle
tunneling which splits the one-electron plane-bands into even and odd with
respect to the mirror-plane between the layers. In this aspect our mechanism
is very different from the interlayer {\em pair}-tunnelling (IPT) mechanism
discussed by Anderson et al.\cite{Science}. Nevertheless, similar to the IPT
model, any attractive interaction between electrons in the same band, such
as the one mediated by even (Raman-active) phonons, enhances $T_c$. This is
opposite to the previously considered single-layer AFSF models in which such
interactions are mutually destructive.

Support for such an enhancement mechanism may be found in the experimental
fact (e.g. Refs.\cite{T*mwave,T*tunn}) that some members of the cuprate
family (Nd$_{2-x}$Ce$_x$CuO$_4$, HgBa$_2$CuO$_4$) behave as conventional $s$%
-wave EP superconductors. MP AFSF theory, on the other hand, would have to
imply principally different mechanisms for this compound and for those with
high $T_c$'s. Another experimental fact which suggests a constructive
interplay between phonon- and non-phonon mechanisms is that in YBa$_2$Cu$_3$O%
$_7,$ the isotope effect increases smoothly when the superconductivity is
suppressed \cite{Frank}. Finally, the most impressive argument is that in
all high-$T_c$ materials $T_c$ is {\it anticorrelated} with the in-plane
antiferromagnetic correlation length $\xi $. In particular, in YBa$_2$Cu$_3$O%
$_{7.0}$, $\xi $ is about one lattice parameter, which would make the
single-layer AFSF superconductivity virtually inoperative. To the contrary,
as we shall argue below, the proposed bilayer model is barely sensitive to
the in-plane AF correlation length $\xi $ at all.

In the following we shall assume a conventional picture in the sense
that the one-electron tunneling between the
planes is allowed both in the normal and in the superconducting states.
This is in contradiction with the IPT
scenario\cite{Science}, but in agreement with some photoemission experiments%
\cite{photo}.
 In this case, the single-particle eigenstates for a bilayer are the
even $|+,{\bf k}\rangle $ and odd $|-,{\bf k}\rangle $ combinations of the
individual plane states and ${\bf k}$ is the 2D Bloch-vector. The properties
of the even and odd bands are discussed in detail in Ref.\cite{TB8}, but,
for the purpose of comparison with the MP model, we use the same band
model as they did.
 We neglect completely
the $k_z$ dispersion due to small intercell $c$-hopping, which can
lead to interesting effects (see, e.g., Ref.\cite{klem}) but
which  are however beyound the scope of
this Letter.
Accordingly, in the following the term ``bands'' always means ``two-dimensional
bands''.
 As regards the interplane hopping inside the unit cell,  $t_{\perp
}\left( {\bf k}\right)$, we assume that it is sufficiently large to set even
 and odd symmetry of the two-dimensional bands, but we neglect, for simplicity,
in the following
numerical calculations the even-odd splitting $
\epsilon _{-}\left( {\bf k}\right) -\epsilon_{+}\left( {\bf k}\right) =
2t_{\perp}({\bf k})$.

The generalization of the MP AFSF model to two-bands is straightforward; one
has only to take into account that the effective vertex for scattering
of an electron from band $i$ to band $j$ by a spin-fluctuation depends on $%
i,j$, while the spectrum of the fluctuations $\chi $ is the same as in MP.
Then, Eqs. (6-8) of MP become%
\begin{eqnarray}
\Sigma _{ij}\left( {\bf k},i\omega _n\right)&=&T\sum_{{\bf q}m}
\sum_{kl}V_{ik,lj} \left(
{\bf k-q},i\omega _n-i\omega _m\right) \nonumber \\
&\times &G_{kl}\left( {\bf q},i\omega _m\right)
\nonumber \\
G_{ij}^{-1}\left( {\bf k},i\omega _n\right)& =&\left[ i\omega
_n-\epsilon \left( {\bf k}\right) +\mu \right] \delta _{ij}-\Sigma
_{ij}\left( {\bf k},i\omega _n\right)\nonumber\\
\Phi _{ij}\left( {\bf k},i\omega _n\right)& =&-T\sum_{{\bf q}m}\sum_{klst}V
_{ik,tj}\left( {\bf k-q},i\omega _n-i\omega _m\right)
\label{MP}\\
\times  &G&_{kl}\left( -{\bf q},-i\omega _m\right)\Phi
_{ls}\left( {\bf q},i\omega _m\right)G_{st}\left( {\bf q},i\omega
_m\right) \nonumber
\end{eqnarray}
where $\Sigma $ and $\Phi $ are respectively the normal and anomalous
self-energies, $G$ is the single-particle Green function, $\epsilon $ the
bare electron energy, and $\sum_{{\bf q}m}$ denotes the {\it average} over
the Brillouin zone plus the sum over the Matsubara frequencies. $V$ is AFSF
pairing interaction, determined by the exchange interaction of electrons
with the AFSF's, $Vij,kl=\int {d{\bf R}d{\bf R}^{\prime }} \sum_{\alpha
\beta \gamma \delta } \langle i\alpha |J({\bf r-R})\sigma _{\alpha \beta
}|j\beta \rangle \\ \times \tilde \chi ({\bf R-R}^{\prime })\langle k\gamma
|J({\bf r-R})\sigma _{\gamma \delta }|l\delta \rangle $ where $J$ is
exchange interaction and $\tilde \chi =\langle {\bf S(R)S(R}^{\prime
})\rangle $ is spin-spin correlation function. For a bilayer, one can let
{\bf R} be a two-dimensional vector and introduce $\tilde \chi =\chi {\bf %
(R-R}^{\prime })I_{uv}$, where $u,v=1,2$ label layers, and $I$ accounts for
interplane correlations, if any. Then the function $\chi $ is the same as in
MP.

The key to the our bilayer AFSF model is the experimental fact that the spin
fluctuations in the bilayer of YBa$_2$Cu$_3$O$_{7-x}$ are always
antiferromagnetically correlated between the planes \cite{tranq,mooK,rossat}.
Even fully oxygenated samples, where the in-plane correlation length is
already of the order of the lattice parameter, show nearly perfect
interlayer correlation\cite{rossat}.
The exchange potential set up by such a spin-fluctuation is therefore {\em %
odd} with respect to the mid-layer mirror plane and, correspondingly,
couples exclusively {\em even and odd} electron states (but neither odd to
odd, nor even to even). In other words, $I_{u\neq v}=-I_{uu}=-1.$ In this
case, after summation over $u,v$ in the expression for $V_{ij,kl}$ and
defining the appropriate coupling constant $g$, Eqs. \ref{MP} become:%
\begin{eqnarray}
\Sigma _{-}\left( {\bf k},i\omega _n\right)&=&Tg^2\sum_{{\bf q}m}\chi \left(
{\bf k-q},i\omega _n-i\omega _m\right)
G_{+}\left( {\bf q},i\omega _m\right) \nonumber
\\
G_{+}^{-1}\left( {\bf k},i\omega _n\right)&=& i\omega
_n-\epsilon \left( {\bf k}\right) +\mu -\Sigma
_{+}\left( {\bf k},i\omega _n\right) \nonumber \\
\Phi _{+}\left( {\bf k},i\omega _n\right)&=&-Tg^2\sum_{{\bf q}m}\chi
\left( {\bf k-q},i\omega _n-i\omega _m\right) \label{MP1}\\
&\times &G_{-}\left( -{\bf q},-i\omega _m\right) \Phi
_{-}\left( {\bf q},i\omega _m\right) G_{-}\left( {\bf q},i\omega
_m\right)  \nonumber \end{eqnarray}
and the same with $+$ and $-$ subscripts interchanged, and $g$ and $\chi $
are the same as in MP.

For reasons of symmetry, the solution of these equations must have the form $%
G_{+}=G_{-}$, $\Phi _{+}=\pm \Phi _{-}$. For the upper choice of the sign,
the Eqs. (\ref{MP1}) reduce precisely to the original MP pairing state. For
the lower choice, the Eqs. (\ref{MP1}) again reduce to the one-plane case,
{\em but now the interaction in the equation for }$\Phi ${\em \ is
effectively attractive}. In other words, now the order parameter has the
opposite sign in the two bands, and therefore the last Eq. in (\ref{MP1})
can be rewritten in terms of $|\Phi |$, and with plus instead of minus on
the right-hand side.

The concept of a superconducting state where two distinctive bands had
the order parameters of the opposite signs was first discussed
in 1973 in connection with semimetals\cite{AS}. More recently,
in a two-layer Hubbard model,
such a solution was found by Bulut et al\cite{Scalap-PRB}
(which they labeled as ``$d_z$'' state)
and in the conventional
superconductivity  theory\cite{we}, where it appears
in case of strongly anisotropic electron-phonon and/or Coulomb
 interaction, or because of a strong interband scattering by magnetic
impurities. In all cases, order
parameter has $s$-symmetry inside each band and changes sign
 between the  bands.

 From Eqs. (\ref{MP1}) it is quite plausible that such an instability is
stronger than the $d_{x^2-y^2}$ one, and will occur at a higher $T_c$. Below
we shall prove this numerically, but before going to numerical results, it
is instructive to get a conceptual understanding about these two different
solutions. The physical reason for having $d$-symmetry in the one-plane
case, is that the AFSF interaction makes pairing energetically favorable
only when it couples parts of the Fermi surface which have opposite signs of
the order parameter\cite{?}. In Y123 the AFSF interaction is peaked at {\bf Q%
}=($\pi /a,\pi /a)$. The shape of the Fermi surface is such that the
condition is satisfied only for $d_{x^2-y^2}$ symmetry. On the other hand,
the small-{\bf q} interaction couples parts of the Fermi surface where the
order parameter has the same sign. This makes pairing unfavorable. Since $%
\chi ({\bf q}\approx {\bf Q})\gg \chi ({\bf q}\approx 0)$, nevertheless,
more is lost by making an $s$-state than by making a $d$-state (which has
been found numerically by MP), because the latter loss is the difference
between the small-$q$ loss and the large-$q$ gain, while in an $s$-state one
loses over the whole Fermi surface\cite{spin-bag}.

Now, coming to the bilayer case, we observe that there is no conflict
between the small and the large $q$'s any more. The AFSF interaction spans
two different sheets of the Fermi surface, which always have order
parameters of the opposite signs. Thus the AFSF interaction is as attractive
for $s$-pairing in a bilayer as it is repulsive in a single plane, and
consequently more attractive than $d$-pairing in a single plane. Of course,
the resulting $s$-state is likely to be highly anisotropic, to take better
advantage of the large $\chi ({\bf q})$ at ${\bf q}\approx {\bf Q}$. This is
similar to Anderson's model\cite{Science}. To demonstrate this effect
numerically, we have solved Eqs.\ref{MP1} with the parameters from MP paper,
and using the same numerical technique. As expected, the maximal eigenvalue
of the last Eq.\ref{MP1} is larger than that for the MP $d$-pairing (about
1.5 compared to 1). Fig.\ref{tc} shows the plot of $T_c$ as a function of
the interaction constant $g$ for both cases. To test the numerics, we have
also solved the original MP equations and obtained the similar results as MP.

 From Fig.\ref{tc}, one immediately observes that the value (0.69 eV) of the
coupling constant $g$ which yields $T_c\approx 90$K for two planes and $s$%
-symmetry, is much smaller than the corresponding value (1.24 eV) for one
plane and $d$-symmetry. Actually, $T_c(s)\sim 2T_c(d)$ for $g$ up to about 1
eV. At stronger couplings, $T_c(d)$ saturates faster than $T_c(s)$ due to
stronger effect of mass renormalization. Similarly, as we shall see below,
the ratio of the maximal gap to $T_c$ tends to be larger for the one-plane
model, for the same $T_c.$

One can also obtain the self-consistent solution for $\Phi $ at $T\ll T_c$.
To do that, one has to include higher-order terms (see, e.g., Ref. \cite
{scal}). In this case one of the Green functions in the Eq.\ref{MP1} for $%
\Phi $ should be replaced by:%
\begin{eqnarray}
\tilde G_-^{-1}({\bf k},i\omega _n)&=&i\omega _n-(\epsilon _{{\bf k}}-\mu
)-\Sigma _-({\bf k},i\omega _n),
\label{MP2}  \\
G_-^{-1}({\bf k},i\omega _n)&=&\tilde G_-^{-1}({\bf k},i\omega _n) \nonumber\\
&-&\Phi _-({\bf k},i\omega _n)\tilde G_-(-{\bf k},-i\omega _n)
\Phi _-(-{\bf q},-i\omega_n)]. \nonumber
\end{eqnarray}
This new set of equations can be solved iteratively, starting with $G=\tilde
G$ (which is correct to first order in $\Phi )$. The actual solution for $%
T=T_c/2$, shown in Fig.\ref{gaps}, was achieved by making two iterations of
Eqs.\ref{MP2}. The frequency-dependent superconducting gap is related to $%
\Phi $ as%
$$
\Delta ({\bf p},i\omega _n)=\frac{\Phi ({\bf p},i\omega _n)}{Z({\bf p}%
,i\omega _n)}=\frac{\Phi ({\bf p},i\omega _n)}{1-Im\Sigma ({\bf p},i\omega
_n)/\omega _n}.
$$
 From Fig.\ref{gaps} we observe that the {\it absolute value }of $\Delta $
behaves similarly in both cases, having a minimum along (11) directions and
a maximum along (10) directions. Furthermore, $|\Delta |$'s in both cases
differ by less than 10\% on two-thirds of the whole Fermi surface, thus
making it extremely difficult to distinguish between the two in an
experiment which does not probe the relative phases of $\Delta .$ In other
words, the order parameter, formally having $s-$symmetry, is still strongly
anisotropic, but nodeless.

Now we shall briefly discuss some experimental consequences of the bilayer
AFSF superconductivity model. It turns out that many difficulties associated
with the original AFSF superconductivity model disappear in the present
version.

(1) In-plane vs. perpendicular-to-the-planes Josephson tunneling. Recent
searches for the $d$-pairing in YBa$_2$Cu$_3$O$_7$ (Refs.\cite{illi,dynes}
and others) still do not give a definite answer. Experiments probing the
angular dependence of the order parameter, as well the existence of the
so-called ``paramagnetic Meissner effect''\cite{par-M}, indicate the
existence of order parameters of opposite signs; Many experiments were
interpreted in terms of $d_{x^2-y^2}$, but such an interpretations can be
questioned because of presence of the chain band\cite{we}.  On the other
hand, the existence of the finite tunneling current perpendicular to the
planes\cite{dynes} is incompatible with $d$-pairing, but compatible with our
model once the simplifying assumption $\epsilon _{-}\left( {\bf k}\right)
=\epsilon _{+}\left( {\bf k}\right) $ is removed\cite{intra}.

(2) The original MP model is very sensitive to the in-plane correlation
length $\xi $: According to Ref.\cite{MP}, $T_c$ drops from 90K to 35K when $%
\xi $ is reduced from 2.3$a$ to $a.$ Experimentally, in fully oxygenated
samples (O$_{6.9-7.0}$) $\xi $ is small and it still decreases closer to the
O$_7$-composition where it becomes less than $a$. $T_c$ is, however, not
sensitive to the oxygen content in this regime. In our model and contrary to
MP, $T_c$ is hardly sensitive to the sharpness of $\chi (q)$, which
predominantly influences the gap anisotropy.

(3) One of the arguments in favor of the AFSF mechanism has been the strong $%
T_c$ suppression upon Zn doping. However, it has remained unclear why chain
disorder hardly affects $T_c.$ (One could argue that the chains are
completely decoupled from the planes, but this is inconsistent with the
strong effect on the Ba $A_{1g}$ phonon mode of the onset of
superconductivity). This finds natural explanation in our AFSF model: The
chain impurity potential is even with respect to the mid-plane reflection
and does therefore not produce scattering between the even and odd bands.

(4) The MP model has difficulty in explaining the continuous change of the
isotope effect with oxygen content, as well as in reconciling the apparent
phonon $s-$wave superconductivity in Nd-cuprate with the assumed AFSF
superconductivity in YBa$_2$Cu$_3$O$_7.$ The basics of this conflict is as
follows: The AFSF interaction is pairing when a pair changes the sign of its
order parameter upon scattering; this is the case MP model for ${\bf q}%
\approx (1,1)\pi /a$. It is {\it de}pairing if there is no sign change. The
opposite is true for the electron-phonon interaction. Obviously, the only
way to reconcile the MP model with the known facts about the role of phonons
in superconductivity is to assume that the electron-phonon interaction,
contrary to the AFSF interaction, is strong for ${\bf q}\rightarrow 0$ and
weak for large $q$'s; this is opposite to common wisdom. In our model the
corresponding assumption is much less painful: One has to assume that the
{\it even} phonons, like for instance $A_{1g}$ Raman-active phonons,
interact with electrons stronger than the {\it odd} ones. This seems quite
plausible and some indirect arguments can be given in support of this
assumption (e.g., the even phonons strongly influence the extended van Hove
singularities\cite{TB8}.

In conclusion, we have shown that from the observed strong antiferromagnetic
correlations between the Cu-O planes in bilayer materials the strongest
superconducting instability due to antiferromagnetic spin fluctuations
appears in the anisotropic $s$-channel, but so that the order parameters in
the bonding (symmetric) and antibonding (antisymmetric) bands  have opposite
signs. This helps to reconcile the magnetic-induced superconductivity model
with many experiments which previously seemed to contradict the magnetic
scenario. In particular, the interrelation between the doping dependencies
of the magnetic and the superconducting properties can be much easier
understood.

We want to thank O. Gunnarsson for many helpful discussions, especially
regarding the relation of this model to the two-band Hubbard model.

\begin{figure}\caption{Critical temperature as a function of the
coupling strength $g$ for a single plane ($d_{x^2-y^2}$)-symmetry) and for
two planes ($p_{z}$-symmetry). Parameters are as in Ref.\protect{\cite{MP}}.
}\label{tc}\end{figure}

\begin{figure}\caption{Superconducting gap $\Delta$
for the lowest Matsubara frequency $\omega _n=\pi k_BT$  in the
2D Brillouin zone
 at $T=0.5T_c$ for a single plane (a), and for
two
planes (b). The Fermi-surface contour shown at the base of the plots
corresponds
to $E_F=-0.37$ eV (0.25 holes/plane). }\label{gaps}\end{figure}

\end{document}